\documentclass[aps,twocolumn,a4paper]{revtex4}

\usepackage{amssymb}
\usepackage{amsmath}
\usepackage{graphicx}
\usepackage{color}

\begin{document}

\title{Anisotropic Magnetoresistance in Manganites: Model and
  Experiment}

\author{Javier D. Fuhr}
\author{Mara Granada}
\author{Laura B. Steren}
\author{Blas Alascio}

\affiliation{Centro At\'omico Bariloche and Instituto Balseiro,
  Comisi\'on Nacional de Energ\'{\i}a At\'omica and Universidad
  Nacional de Cuyo, 8400 San Carlos de Bariloche, Argentina.}

\begin{abstract}
  We present measurements of anisotropic magnetoresistance of
  La$_{0.75}$Sr$_{0.25}$MnO$_3$ films deposited on (001) SrTiO$_3$
  substrates, and develop a model to describe the low temperature AMR
  in manganites. We measure an AMR of the order of 10$^{-3}$ for the
  current $I$ parallel to the [100] axis of the crystal and vanishing
  AMR for $I//$[110], in agreement with the model predictions.
\end{abstract}

\date{\today}
\maketitle

% ****************************************************

{\it Introduction} - 
Colossal magnetoresistant manganites have been investigated thoroughly
since the discovery of their magnetoresistive properties.
\cite{Jin94,Hwang96,Coey99,Dagotto01,Israel07} Anisotropic
magnetoresitance (AMR) of these compounds has also been investigated
since it may give rise to application of these materials in
electronics.\cite{Ziese98, Bason04, Bibes05, Infante06, Yau07} It has
been found that the resistivity of polycrystalline ferromagnetic
metals and alloys, in the magnetic ordered state, depends on the angle
$\theta$ between the magnetization $M$ and the electric current
$I$. This dependence has the form\cite{Campbell2}
\begin{equation}  \label{eq:AMR}
\rho(\theta) = \frac{\rho_{\parallel}+ 2\rho_{\perp}}{3}+
\left(\cos^2 \theta - \frac{1}{3}\right)(\rho_{\parallel} -
\rho_{\perp}),
\end{equation}
being $\rho_{\parallel}$ and $\rho_{\perp}$ the resistivity measured
with current flowing parallel or perpendicular to the magnetization,
respectively. The AMR is defined as
\begin{equation}
\rho _{A}=\frac{\rho _{\parallel }-\rho _{\perp }}{\frac{1}{3}\rho
_{\parallel }+\frac{2}{3}\rho _{\perp }}.  \label{eq.AMRFert}
\end{equation}

The parameters obtained by fitting the experimental results with
Eq.~(\ref{eq:AMR}) are used to calculate $\rho_A$ from
Eq.~(\ref{eq.AMRFert}), which quantifies the anisotropy of the
resistivity.

AMR has been observed in conventional metallic systems and in colossal
magnetoresistant materials. However, their sign and temperature
dependences are quite different: while in most conventional metals
$\rho _{A}$ is positive\cite{Campbell} and decreases with decreasing
magnetization or increasing temperature, in manganites it is an order
of magnitude lower, of opposite sign, and its temperature dependence
is non monotonic.\cite{Ziese98, Bason04, Bibes05, Infante06, Yau07}
These differences point to the fact that different mechanisms must be
in action in the different materials. The model proposed by Campbell
{\it et al.}\cite{Campbell} based on the scattering of $s$ waves on
the $d$ sites of the material has been successful to the understanding
of AMR in metallic alloys (describing properly the effects of impurity
concentration and temperature dependence) and it has also been
mentioned in reference to measurements in manganites. However, it is
not appropriate to apply this model to manganites where the carriers
(electrons or polarons) move by hopping between the $d$ states of the
transition metal. We present here measurements on manganite films and
a model that describes the resulting anisotropy as well as its
dependence on the direction of the current to the crystalline axes.

% ****************************************************

{\it Experiment} - 
The ferromagnetic manganite La$_{0.75}$Sr$_{0.25}$MnO$_3$ (LSMO)
presents a nearly cubic perovskite structure\cite{Urushibara95} that
makes this system appropriate to compare with the model, which is
developed for a cubic lattice of Mn ions. The samples were deposited
on (001) SrTiO$_3$ substrates by dc sputtering.  The films grow
textured following the (001) orientation of the substrate, as
confirmed by X-ray diffraction.\cite{Mara1} This substrate, having a
lattice constant similar to that of LSMO, induces little distortion on
the film compared to the bulk manganite.

We present electrical transport measurements performed on films of
different thicknesses, using the longitudinal four-lead configuration
with the electrical contacts on the plane of the films. The magnetic
field was applied parallel to the plane of the samples with an
electromagnet mounted on a rotating platform. All the measurements
were carried out with an applied field $H = 10$ kOe, which is strong
enough to saturate the magnetization. So the angle $\theta$ between
the electrical current and the magnetic field is assumed to be the
same as the angle between the current $I$ and the magnetization $M$.

\begin{figure}[ht]
\begin{center}
\includegraphics[width=8cm]{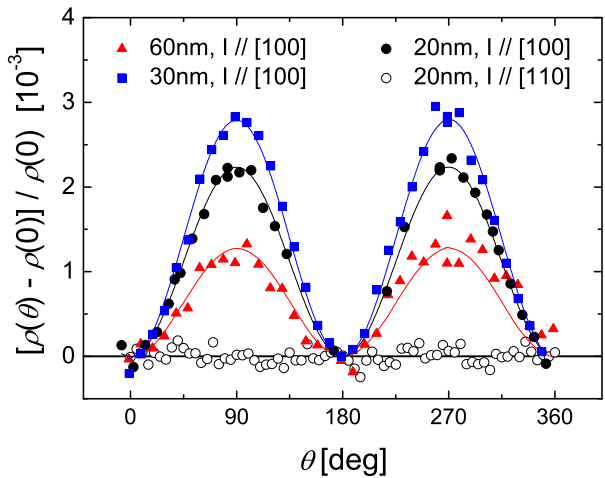}
\end{center}
\caption{(Color online) Normalized resistivity measured at $T = 88$ K
  with the current $I$ applied parallel to the crystalline direction
  [100] for films of three different thicknesses (20 nm, 30 nm and 60
  nm). Hollow circles represent the resistivity measurements with
  $I//$[110] for the 20 nm sample.  $\theta$ is the angle measured
  between the direction of the electrical contacts and the applied
  magnetic field ($H = 10$ kOe). A fit with Eq.~(\ref{eq:AMR}) is also
  shown (lines).}
\label{mediciones}
\end{figure}

Figure~\ref{mediciones} shows the normalized resistivity $
[\rho(\theta)-\rho(0)]/\rho(0)$ measured at $T = 88$ K with the
current applied parallel to the crystalline direction [100] showing
agreement with Eq.~(\ref{eq:AMR}) and previous AMR
measurements.\cite{Infante06, Yau07} We have not found any systematic
dependence on the thickness of the films. Prompted by the model
prediction of vanishing AMR we also performed measurements with the
current parallel to [110] direction. The resulting AMR is displayed in
Fig.~\ref{mediciones} by empty circles, where we see negligible
dependence of the resistivity with the magnetic field direction.
The general formula obtained by D\"oring\cite{Doring,Campbell2} for
the resistivity of a cubic ferromagnet with the magnetization in the
($\alpha_1$, $\alpha_2$, $\alpha_3$) direction and the current in the
($\beta_1$, $\beta_2$, $\beta_3$) direction is
\begin{eqnarray}
  \label{eq:Doring}
  \rho &=& \rho_0 [1+k_1 (\alpha_1^2\beta_1^2 +
  \alpha_2^2\beta_2^2 + \alpha_3^2\beta_3^2 - {1 \over 3}) 
  \nonumber \\ 
  && + 2 k_2 (\alpha_1 \alpha_2 \beta_1 \beta_2 + \alpha_2 \alpha_3
  \beta_2 \beta_3 + \alpha_3 \alpha_1 \beta_3 \beta_1) \nonumber \\
  && + k_3 (s - {1\over 3}) + k_4 (\alpha_1^4 \beta_1^2 + \alpha_2^4
  \beta_2^2 + \alpha_3^4 \beta_3^2 + {2\over 3} s - {1\over 3})
  \nonumber \\
  && + 2 k_5 (\alpha_1 \alpha_2 \alpha_3^2 \beta_1 \beta_2 + \alpha_2
  \alpha_3 \alpha_1^2 \beta_2 \beta_3 + \alpha_3 \alpha_1 \alpha_2^2
  \beta_3 \beta_1 )] 
\end{eqnarray}
where $s=\alpha_1^2\alpha_2^2 + \alpha_2^2\alpha_3^2 +
\alpha_3^2\alpha_1^2$. Considering the geometry studied here, we
obtain
\begin{equation}
  \label{eq:DoringSimp}
  \frac {\Delta\rho(\theta)}{\rho(0)} \propto  
  \begin{cases}
    C_1 \cos^2(\theta) + C_3 [\cos^4(\theta) -
    \cos^2(\theta) ]& \text{for $I//$[100]} \\
    C_2  \cos^2(\theta) - C_3 [\cos^4(\theta) -
    \cos^2(\theta) ] & \text{for $I//$[110]}
  \end{cases}
\end{equation}
where $C_1 = k_1+k_4$, $C_2 = k_2$ and $C_3 = (k_4-3 k_3)/3$.  The
experimental results imply that the constants $C_2$ and $C_3$ are
almost zero, and the only finite contant is $C_1 = k_1+k_4$.

% ****************************************************

{\it Model} -
In La$_{1-x}$Sr$_{x}$MnO$_{3}$, the Mn$^{3+/4+}$ ions form a nearly
simple cubic lattice, with oxygen ions located between each pair of Mn
neighbors and La/Sr ions at the body center of the cube. The
octahedral symmetry around each Mn splits the $3d$ levels into a lower
energy $t_{2g}$ triplet and a higher energy $e_{g}$ doublet.  Due to
Hund's rule, the three $t_{2g}$ orbitals are all singly occupied with
their spins coupled to form a total spin $S=3/2$. The additional
electron on a Mn$^{3+}$ ion occupies the $e_{g}$ orbitals and it is
considered, due again to the strength of the exchange term, to align
its spin parallel to the $t_{2g}$ electrons. We model then the $e_{g}$
electrons by a spinless Hamiltonian on a cubic lattice:
\[
H=\sum_{\left\langle ij\right\rangle \alpha \beta }t_{ij}^{\alpha \beta
}c_{i\alpha }^{\dagger }c_{j\beta }
\]
with $t_{ij}^{\alpha \beta }$ the hopping integrals that depend both
on the type of orbitals $\alpha ,\beta $ and on the direction between
neighbouring sites $i,j$.\cite{PRL} At low temperatures we assume that
the localized spins of the $t_{2g}$ electrons are all aligned with the
external magnetic field.

When spin-orbit (SO) coupling is included, the degeneracy of the
$e_{g}$ orbitals ($|z\rangle = |3z^2-r^2\rangle$, $|x\rangle =
|x^2-y^2\rangle$) is lifted. By symmetry, there is no coupling between
$e_{g}\uparrow$ and $e_{g}\downarrow$ orbitals. Moreover, we take into
account only the coupling between the $e_{g}\uparrow$ and
$t_{2g}\uparrow$ orbitals (separated by the crystal field by $\sim
1.5$eV), and neglect the coupling with the $t_{2g}\downarrow$ orbitals
(separated by $\sim 6$eV\cite{Saitoh95}).  The character of the two,
now not degenerated, orbitals ($|1\rangle $, $|2\rangle $) depends on
the direction of the magnetic field. From second order perturbation
theory, the shift and coupling of the two original $e_g$ orbitals for
the magnetization in a given direction ($\theta_B$,~$\phi_B$) are
\begin{equation}
  \label{eq:H1}
  H_1 = g \left (
    \begin{array}{cc}
      3\sin^2(\theta_B) & \sqrt{3}\sin^2(\theta_B)\cos(2\phi_B) \\
      \sqrt{3}\sin^2(\theta_B)\cos(2\phi_B) &
      \sin^2(\theta_B)+4\cos^2(\theta_B)
    \end{array}
    \right )
\end{equation}
where $g=\lambda^2/\Delta_{CF}$, being $\lambda$ the SO coupling
constant and $\Delta_{CF}$ the crystal field splitting between
$t_{2g}$ and $e_g$ orbitals. From this perturbation we obtain the new
energy levels ($\varepsilon_{1,2} = g\,(2\mp\Delta)$) and the
corresponding eigenvectors
\begin{equation}
  \label{eq:eigensystem}
  |1\rangle,|2\rangle = \frac{(a\mp\Delta)}{r_{1,2}} | z
  \rangle + \frac{b}{r_{1,2}} |x\rangle
\end{equation}
where
\begin{eqnarray}
\nonumber a & = & \sin^2(\theta_B)-2\cos^2(\theta_B)\\
\nonumber b & = & \sqrt{3}\sin^2(\theta_B)\cos(2\phi_B)\\
\nonumber \Delta & = & \sqrt{a^2+b^2}\\
\nonumber r_{1,2} & = & \sqrt{(a\mp\Delta)^2+b^2}.
\end{eqnarray}

Assuming an isotropic relaxation time $\tau $, we calculate the
conductivity in a given direction $\hat{r}$ by\cite{conduct}
\begin{equation}
  \sigma_{\hat{r}}=e^2\tau \int
  d^{3}k\,|v_{\hat{r}}(\vec{k})|^{2}\,
  \frac{\partial f}{\partial \varepsilon (\vec{k})}
  \label{eq:conduct}
\end{equation}
with $v_{\hat{r}}(\vec{k}) =
\hat{r}\cdot\vec{\nabla}\varepsilon(\vec{k}) $ and $ f(\varepsilon)$
the Fermi function.

We have taken as the energy reference the hopping $t$ between two
$|z\rangle$ orbitals in the $\hat{z}$ direction. Therefore, the only
parameter in the model is the constant $g/t=\lambda^2/(\Delta_{CF}\,
t)$. We take $g/t=0.001$ to fit the experimental results for the 30 nm
film. This value is perfectly consistent with the atomic value of the
LS coupling ($\lambda=0.04$eV), the crystal field splitting
($\Delta_{CF}=1.5$eV) and a hopping of $t=0.4$eV.\cite{Ziese98}

% ****************************************************

\begin{figure}[ht]
\begin{center}
\includegraphics[width=8cm]{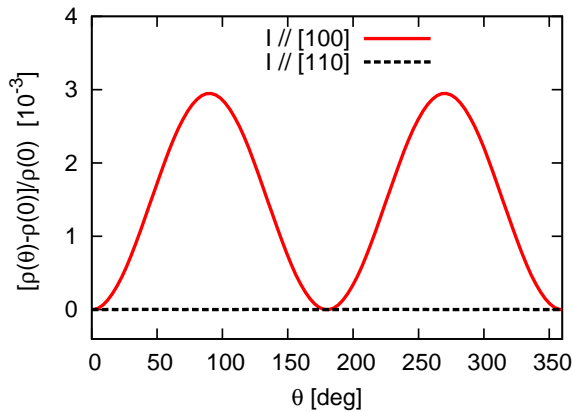}
\end{center}
\caption{(Color online) Calculated normalized resistivity in the [100]
  and the [110] directions, as a function of the magnetic field
  direction $\theta$ measured from the corresponding current
  direction. The magnetic field rotates in the $\hat{x}$-$\hat{y}$
  plane.}
\label{fig:AMR_n001}
\end{figure}

We first consider the case with the magnetic field rotating in the
$\hat{x}$-$\hat{y}$ plane which corresponds to films with normal
[001]. We calculate the conductivity in two directions ([100] and
[110]) and obtain the dependence of the normalized resistivities
$[\rho(\theta)-\rho(0)]/\rho(0)$ with the direction of the magnetic
field. We show the calculated AMR in Fig.~\ref{fig:AMR_n001}, where
the angle $\theta$ of the magnetic field is measured from the
corresponding current direction. Although these calculations were
carried out numerically, the result for $I//$[100] is
indistinguishable from a $\cos ^2 \theta$ dependence like that of
Eq.~(\ref{eq:AMR}). We can see that, in agreement with the
experimental results shown in Fig.~\ref{mediciones}, the resistivity
for $I//$[110] does not depend on the direction of the magnetic field.
This result can be understood from the $\cos^2(\theta)$ dependence of
the resistivity, the cubic symmetry, and the dependence of
conductivity with current direction given by
Eq.~(\ref{eq:conduct}). In fact, from the cubic symmetry we have that
$\sigma_{[010]}(\theta) = \sigma_{[100]}(\theta-\pi/2)$, and from
Eq.~(\ref{eq:conduct}), we have $\sigma_{[110]}(\theta) =
[\sigma_{[100]}(\theta) + \sigma_{[010]}(\theta) ]/2$. Then, assuming
$ \sigma_{[100]}(\theta) = \overline{\sigma} + \Delta\sigma
[\cos^2(\theta)-1/2] $ we obtain $\sigma_{[110]}(\theta) =
\overline{\sigma}$, which clearly does not depend on the magnetic
field direction $\theta$.

\begin{figure}[ht]
\begin{center}
\includegraphics[width=8cm]{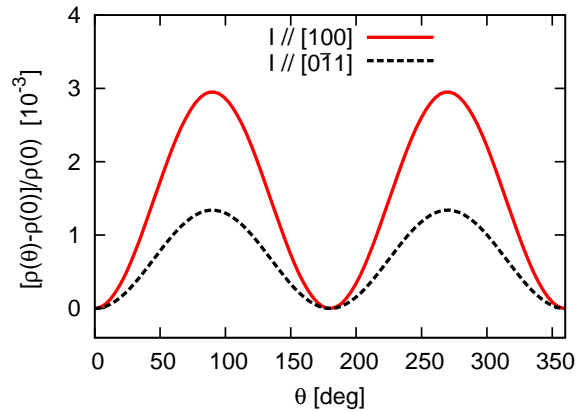}
\end{center}
\caption{(Color online) Calculated normalized resistivity in the [100]
  and the [0$\bar{1}$1] directions, as a function of the magnetic
  field direction $\theta$ measured from the corresponding current
  direction. The magnetic field rotates in the [100]-[0$\bar{1}$1]
  plane.}
\label{fig:AMR_n011}
\end{figure}

We now consider another case that corresponds to films with normal
[011]. The magnetic field now rotates in the [100]-[0$\bar{1}$1] plane
and we calculate the conductivity in the two nonequivalent directions
[100] and [0$\bar{1}$1]. We show these calculated AMR in
Fig.~\ref{fig:AMR_n011}, where as before the angle $\theta$ of the
magnetic field is measured from the current direction. It is difficult
to compare this prediction with previous experimental
results,\cite{Infante06,Bibes05} since the behaviour of AMR strongly
depends on the samples.

\begin{figure}[ht]
\begin{center}
\includegraphics[width=8cm]{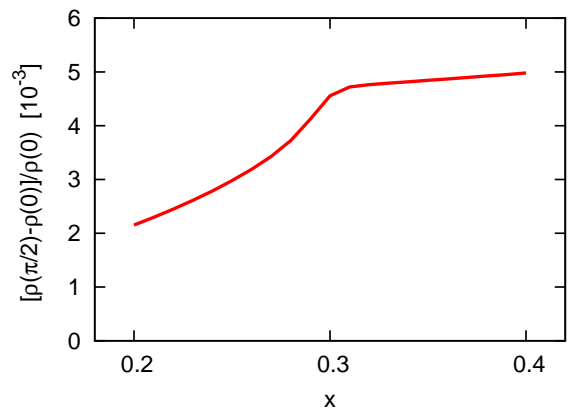}
\end{center}
\caption{(Color online) Dependence of the AMR with the doping $x$.}
\label{fig:AMR_nel}
\end{figure}

In Fig.~\ref{fig:AMR_nel}, we show the variation with doping of the
AMR defined as $[\rho(\pi/2)-\rho(0)]/\rho(0)$ for the first case
(Fig.~\ref{fig:AMR_n001}), and the current in the [100] direction. We
can see a strong increasing of the AMR with the doping, in particular
for $x \lesssim 0.3$. For these calculations we keep all parameters
fixed, except for the occupation number $n=1-x$. One can expect that
both the hopping $t$ and the crystal field splitting $\Delta_{CF}$
will increase with the doping. As a consequence the constant $g/t$,
and therefore the AMR, will decrease.  Moreover, we have not taken
into account neither the effect of the substrate on the film nor the
deviation from the cubic symmetry observed in manganites by changing
the doping.\cite{Urushibara95}

{\it Conclusions} -
We have measured the low temperature anisotropic magnetoresistance of
(001) La$_{0.75}$Sr$ _{0.25}$MnO$_{3}$ films. We have also formulated
a simple model to calculate its angular dependence for different
directions of the current to the crystalline axes. The model explains
satisfactorily the sign and magnitude of the measured AMR for the
current along the [100] axis. Furthermore, it also accounts for the
vanishing of the AMR when measured with the current flowing in the
[110] direction.  More research is necessary to extend the model to
describe the dependence of the AMR with doping, temperature, and
orientation and matching with substrate.

This work was partially supported by ANPCyT (PICT 03-12397, PICT
33304, PICT 03-12742, PICT 05-38387), U.N. Cuyo and CONICET (PIP 5250,
PIP 5342-05). J.D.F, L.B.S. and B.A. are members of CONICET,
Argentina. M.G. acknowledges a fellowship from CONICET. L.B.S. is a
fellow of the Guggenheim Foundation.

\end{document}